\documentclass[sn-basic]{sn-jnl}
 
\usepackage{graphicx}%
\usepackage{multirow}%
\usepackage{amsmath,amssymb,amsfonts}%
\usepackage{amsthm}%
\usepackage{mathrsfs}%
\usepackage[title]{appendix}%
\usepackage{xcolor}%
\usepackage{textcomp}%
\usepackage{manyfoot}%
\usepackage{booktabs}%
\usepackage{algorithm}%
\usepackage{algorithmicx}%
\usepackage{algpseudocode}%
\usepackage{listings}%
\usepackage{pifont} 

\newcommand{\cmark}{\ding{51}} 
\newcommand{\xmark}{\ding{55}} 
\newcommand{\umark}{$\bigcirc$} 

\theoremstyle{thmstyleone}%
\theoremstyle{thmstyletwo}%

\theoremstyle{thmstylethree}%

\begin{document}

\title[Towards Automated Formal Verification of Backend Systems with LLMs]{Towards Automated Formal Verification of Backend Systems with LLMs}








\author[1]{\fnm{Kangping} \sur{Xu}}\email{xkp24@mails.tsinghua.edu.cn}

\author[1]{\fnm{Yifan} \sur{Luo}}\email{luoyf24@mails.tsinghua.edu.cn}

\author*[1,2]{\fnm{Yang} \sur{Yuan}}\email{yuanyang@tsinghua.edu.cn}

\author*[1,2]{\fnm{Andrew Chi-Chih} \sur{Yao}}\email{andrewcyao@tsinghua.edu.cn}

\affil*[1]{\orgdiv{Institute for Interdisciplinary Information Sciences}, \orgname{Tsinghua University}, \orgaddress{\city{Beijing}, \country{China}}}

\affil[2]{\orgname{Shanghai Qizhi Institute}, \orgaddress{\city{Shanghai}, \country{China}}}


\abstract{
Software testing plays a critical role in ensuring that systems behave as intended. However, existing automated testing approaches struggle to match the capabilities of human engineers due to key limitations such as test locality, lack of general reliability, and business logic blindness. In this work, we propose a novel framework that leverages functional programming and type systems to translate Scala backend code into formal Lean representations. Our pipeline automatically generates theorems that specify the intended behavior of APIs and database operations, and uses LLM-based provers to verify them. When a theorem is proved, the corresponding logic is guaranteed to be correct and no further testing is needed. If the negation of a theorem is proved instead, it confirms a bug. In cases where neither can be proved, human intervention is required. We evaluate our method on realistic backend systems and find that it can formally verify over 50\% of the test requirements, which suggests that half of a testing engineer's workload can be automated. Additionally, with an average cost of only \$2.19 per API, LLM-based verification is significantly more cost-effective than manual testing and can be scaled easily through parallel execution. Our results indicate a promising direction for scalable, AI-powered software testing, with the potential to greatly improve engineering productivity as models continue to advance.
}




\maketitle

\begin{figure}[!htbp]
    \centering
    \includegraphics[width=\textwidth]{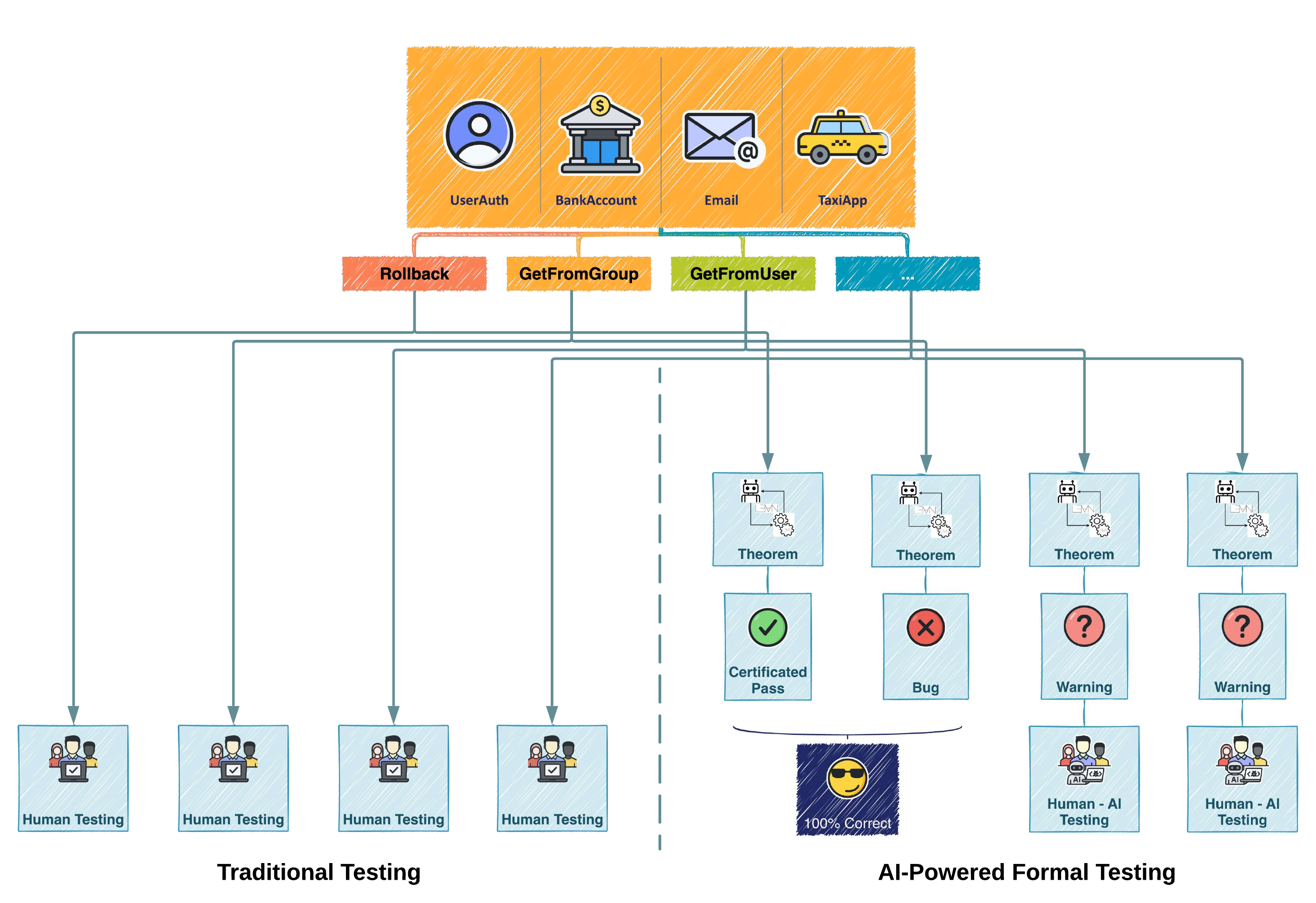} 
    \caption{\textbf{Comparison between traditional human testing and our new framework.} Traditional human testing (left) requires extensive labor and relies on manually written test cases. Our framework (right) addresses this limitation by introducing an LLM-based formalizer that automatically translates backend implementations and documentation into Lean formal system, with another LLM-based prover for proving these theorems. Experimental results demonstrate that our approach can formally verify over 50\% of API specifications and table properties, which significantly reduces manual testing workload.}
    \label{fig:enter-label}
\end{figure}

\section{Introduction}
\label{sec:introduction}

Today, the global software testing industry generates over \$55 billion annually~\citep{researchnester2025software}, reflecting its crucial role in software engineering. In the context of AI-driven development, testing is becoming even more important: as LLMs can now generate massive code rapidly,   the role of software engineers is shifting from coding to verifying AI-generated outputs. A natural question arises: can we automate this verification process?

Extensive and intriguing research efforts have been directed toward this goal, 
from static analyzers and linters to unit testing frameworks~\citep{googletest2025,pylint2025,sonarqube2025,chen2024chatunitest,lops2024system,nama2024artificial,tihanyi2025vulnerability,yang2024formal}. However, to the best of our knowledge, existing approaches still fall short of replicating the work of real-world testing engineers. In particular, they suffer from three main limitations:
\begin{enumerate}
    \item \textbf{Test Locality:} they generally focus on testing individual functions or single files, rather than taking entire systems into consideration.
    \item \textbf{Lack of General Reliability:} they either require language-dependent and predefined verification patterns, or rely on LLMs for automatic test that reduces the reliability of verification.
    \item \textbf{Business Logic Blindness:} they often overlook actual business logic, concentrating primarily on aspects like memory management and security vulnerabilities.
\end{enumerate}

Yet, these are precisely the aspects that real-world test engineers care about most. They require methods capable of testing entire systems, ensuring verification accuracy, and aligning well with business logic. To address this gap, we must ask: is there a framework that can simultaneously meet these requirements?

Topos theory~\citep{johnstone2002sketches} serves as a principled foundation to this end—a powerful mathematical framework that decomposes complex structures into well-organized, composable components. While the theory itself is abstract, we use its central insight: modeling business logic as a type system and applying formal reasoning to validate it. Through a functional and compositional design, this approach enables scalable and rigorous testing, even for complex real-world systems.

More specifically, our method is rooted in functional programming~\citep{hughes1989functional}, primarily utilizing Scala~\citep{odersky2008programming} due to its expressive type system. Scala allows us to express all logic through pure functions, thereby enjoying several desirable properties. For instance, any complex function can be decomposed into a composition of simpler ones. Furthermore, the correctness of each function is independent of external environmental factors. This compositional property directly addresses the issue of Test Locality, as it allows us to verify each small function in isolation while still ensuring the correctness of the complete system. Moreover, because Scala runs on the JVM and is interoperable with Java, it can leverage the vast ecosystem of Java libraries and perform nearly everything that Java can do.

However, basic functions alone can only express low-level data transformations and are insufficient to capture the full semantics of business logic. To address this limitation, we incorporate type systems to structurally encode key aspects of the business domain—such as entities, constraints, and process flows -- within Scala's functional programming paradigm. While actual business operations often require reasoning over runtime values, the type system provides a formal scaffold that constrains and guides how such operations are composed and executed.

Building on this foundation, we use the Lean formal proof system ~\citep{moura2021lean} to enable automated verification. Since our backend is already structured with a Scala type system, translating it into Lean is straightforward, as they share similar expression styles. We then use LLMs to generate properties the system should satisfy, express them in Lean, and attempt automatic proofs. If a proof succeeds, we can conclude that the code is formally correct. If the proof fails, it may indicate either a bug in the code or that the model lacks the capability to complete the proof. To tell these cases apart, we try proving the negation of the original statement. If the model can prove the negation, it confirms a bug; if not, the failure is likely due to the model’s limitations, in which case human intervention is required.
Our approach offers a new direction for general verification, as it is language-agnostic, applies to all parts of the system, and enables formal reliability when proofs succeed.

Based on our experimental results, our method can formally verify over 50\% of test requirements in real-world backend systems, suggesting that more than half of a testing engineer’s workload could be automated. It costs only \$2.19 per API on average---lower than manual testing---while providing stronger guarantees through formal proofs. For the verified portion, correctness is mathematically ensured, offering higher assurance than traditional testing. Additionally, our approach detects bugs in faulty systems by generating concrete counterexamples with over 70\% success, allowing engineers to focus on a smaller set of remaining issues and accelerating the debugging process. Thanks to the composable structure of our functional programming framework, the method scales efficiently to large systems through parallel LLM execution, maintaining both speed and accuracy.

These contributions establish a practical framework for software-level verification of complete backend systems, overcoming the limitations of previous approaches focused on individual components. By effectively combining LLMs with formal methods at an architectural scale, we demonstrate the potential for AI-enhanced verification. We hope that this will be a starting point for deploying these methods at production scale and improving verification capabilities for complex system behaviors.


\section{Related Work}
\label{sec:related_work}

\subsection{Program Verification Tools}
\label{subsec:program_verification_tools}

Formal verification of software and hardware systems has been extensively
studied, with various tools developed for different verification needs.
Automated tools like the Efficient SMT-based Bounded Model Checker (ESBMC)~\citep{esbmc2024},
Z3~\citep{de2008z3}, and CVC4~\citep{barrett2011cvc4} are widely used for detecting
vulnerabilities through bounded model checking. For more rigorous verification, interactive
theorem provers such as Coq, Isabelle, and Lean provide powerful frameworks,
often enhanced by automated tactics like CoqHammer~\citep{czajka2018hammert} and
TacTok~\citep{first2020tactok}. However, these systems require manual
translation of source code into their formal languages, which remains a significant
hurdle.

To reduce this burden, several languages and tools have integrated verification directly into the programming workflow. Dafny~\citep{leino2010dafny} and Verus~\citep{lattuada2023verus} (for Rust) are designed with built-in verification support, while libraries like Stainless~\citep{stainless} extend general-purpose languages (e.g., Scala) with verification capabilities built upon SMT-solvers. Though equipped with verification support, human efforts are still needed to provide the specifications at the function level. The Soda language~\citep{mendez2023soda} takes this further by compiling
Scala code to both executable programs and verifiable Lean projects, though
complex proofs still require human intervention without LLM's support.

\subsection{Formal Verification Practices}
Researchers have invested significant effort in verifying existing software and hardware designs using formal tools. \citet{yang2024formal} provides an overview of formally verified projects, including microprocessor designs \citep{goel2022microprocessor}, network protocols \citep{zheng2023automated}, C compilers \citep{leroy2016compcert}, file systems \citep{chajed2022verifying}, VLSI hardware designs \citep{seligman2023formal}, and operating system kernels \citep{klein2009sel4}. These projects typically require years of expert work to verify.

Despite some automation, most verification practices remain labor-intensive, limiting the adoption of formal verification in real-world projects. This challenge has prompted growing interest in using deep learning techniques to further automate the verification process.

\subsection{LLM-Driven Code Verification}
\label{subsec:llm_driven_code_verification}

Recent work has explored using large language models (LLMs) to automate parts of
the formal verification pipeline. A common focus is auto-formalization—translating
code or specifications into formal representations—where techniques like
Retrieval-Augmented Generation (RAG) have improved accuracy and reduced
hallucinations~\citep{bhatia2024verified, liu2025rethinking}. LLMs also show
promise in generating program specifications, invariants, and theorem statements~\citep{tihanyi2023new,cao2025informal},
which are critical for verification but traditionally require expert effort.

However, most existing approaches still rely on rule-based components for core verification
tasks. For example, FVEL~\citep{lin2024fvel} uses LLMs for theorem generation
but depends on rule-based translation to Isabelle for formalization. Other work~\citep{wu2023lemur,mugnier2024laurel,si2020code2inv,misu2024towards,loughridge2024dafnybench}
leverages verification-friendly languages (e.g., Dafny), where LLMs insert assertions
that are checked by traditional solvers. While the finetuning approach and
thorough evaluations in~\citep{cao2025informal} demonstrate LLMs' potential on isolated
verification tasks, no existing pipeline fully integrates LLMs across the entire
process—from formalization to proof generation—for complex, real-world systems.

\subsection{Domain-Specific Verification Applications with AI}
\label{subsec:domain_specific_verification_applications}

Formal verification has been successfully adapted to various specialized domains,
each with distinct requirements. In hardware verification, tools like~\citep{gadde2024all}
employ formal checkers for SystemVerilog, while other approaches~\citep{liu2024domain}
train domain-specific LLMs to assist in VLSI design and verification. For smart contracts,
PropertyGPT~\citep{liu2024propertygpt} leverages LLMs to generate and verify safety
properties through automated provers. Additional applications include memory
safety verification~\citep{lattuada2023verus} (e.g., for Rust programs) and formal analysis of
cryptographic protocols~\citep{curaba2024cryptoformalevalintegratingllmsformal}, demonstrating the versatility of verification techniques
across different security-critical domains.

Despite these advances, existing approaches and also evaluation benchmarks~\citep{si2020code2inv,beyer2024state} focus on isolated programs or
predefined low-level properties (e.g., memory safety). They do not address the challenges
of verifying large, interconnected systems, where compositional
reasoning and flexible natural-language specifications of the services are essential.
Our work fills this gap by introducing an LLM-driven pipeline capable of end-to-end
verification for complex, real-world backend architectures.

\section{Preliminaries}

\subsection{Functional Programming}
\label{sec:fp_foundation}

Functional programming is a programming paradigm centered around the evaluation of pure mathematical functions, which can be represented as:
\[
y = f(x), \quad f: X \rightarrow Y 
\]
Here, the function \( f \) takes an input \( x \in X \) and deterministically produces an output \( y \in Y \). A function \( f \) is said to be pure if it adheres strictly to the following properties:
\begin{itemize}
    \item \textbf{Referential transparency:} The output of the function depends solely on its given input and not on any external state or hidden variables. Thus, calling the function with the same input always yields the same output.
    \item \textbf{No side-effects:} Evaluation of the function does not affect the outside world, i.e., it does not modify global state, I/O operations, or mutable data structures.
\end{itemize}

This paradigm differs significantly from imperative programming languages (e.g., C, Python), which commonly allow and often encourage mutable states, side effects, and external dependencies such as global variables, pointers, or object mutations. Scala, as a hybrid language, effectively supports functional programming by enabling pure function implementations and immutable data structures through libraries like Cats and Cats Effect.

One key advantage of functional programming is its capacity for modularization. Complex computations can be decomposed into smaller, simpler, and easily testable pure functions. These smaller functions can be composed mathematically, greatly simplifying debugging, testing, and formal verification.

Functional languages and their implementations become practically applicable through the introduction of types. From the perspective of type theory, types serve as formal abstractions that classify program constructs and specify constraints on their composition. Types ensure the correctness of computations by enforcing strict contracts on function inputs and outputs. Basic types (primitive types) such as \texttt{Int}, \texttt{Bool}, \texttt{String}, and \texttt{Enum} serve as the foundational elements. More advanced types can then be constructed using these primitives, including product types (e.g., tuples or records), sum types, and higher-order types. Such types form robust data structures capable of precisely capturing complex business logic.

\textbf{Type-driven development} in functional programming involves designing applications by first defining precise, domain-specific types, subsequently implementing functions that transform these types. In this paradigm, the types themselves document and constrain the application's logic, resulting in code that is robust, expressive, and self-explanatory.

\subsection{Monad}
\label{sec:monad}
While pure functional programming has many theoretical and practical advantages, it inherently avoids side effects, which are essential for real-world applications such as I/O operations, database interactions, or network communications. To reconcile purity with the need for side effects, functional programming employs the concept of a \textit{Monad}—a structure borrowed from category theory that encapsulates computations along with their context or side effects.

Intuitively, a monad can be viewed as a computational context or a ``box'' that encapsulates potential side effects. It preserves purity by deferring side-effectful computations as first-class values that can be manipulated without actually executing the side effects until explicitly instructed to do so. 

Monads have two essential properties:
\begin{itemize}
    \item \textbf{Encapsulation of Side-effects:} Side-effectful operations  are encapsulated as pure computations inside monads. The monad itself remains pure until explicitly ``activated'' by an external interpreter or runtime.
    \item \textbf{Composability:} Monads enable the compositional chaining of operations. Multiple monadic computations can be combined to form larger computations, maintaining the purity and modularity of code.
\end{itemize}

For instance, in Scala, the library Cats Effect provides an \texttt{IO Monad}, facilitating pure functional programming while enabling real-world side effects. Programmers write purely functional code to describe computations within the \texttt{IO} context, and side effects are deferred until explicitly executed by the framework. Thus, the program remains referentially transparent and pure, while still interacting effectively with the external environment.

\subsection{Algebraic Effects and Handlers}
\label{sec:algebraic_effects_and_handlers}
While monads work well for managing side effects in functional programming, they sometimes hide where exceptions come from, especially when using multiple APIs together. This creates challenges when trying to verify code formally in systems like Lean. To solve this problem, we draw inspiration from the strategy of \texttt{Algebraic Effects}~\citep{plotkin2009handlers, pretnar2015introduction, bauer2015programming}, which clearly list all possible outcomes of operations, making our verification process more straightforward.

Algebraic effects handle side effects by treating operations like data changes, errors, and external interactions as separate elements. Unlike monads that combine effects with code, algebraic effects separate what an effect does from how it's handled, making code more flexible since different program parts can handle effects in various ways. Several languages have embraced this approach and handled effects as first-class citizens, including Eff~\citep{kiselyov2018eff}, Koka~\citep{leijen2014koka}, and Frank~\citep{lindley2017}.

This design offers key advantages for our formalization: by explicitly defining all possible outcomes (both successes and errors) and maintaining clear traceability of effect origins, we ensure our formal verification can handle every scenario while preventing important exceptions from being hidden when APIs interact with each other.

\subsection{Backend System Structure}
\label{sec:backend_system_structure}
To effectively realize the benefits of functional programming within practical backend systems, we adopt a microservice-oriented architecture, decomposing the backend system into modular, loosely-coupled components, each of which encapsulates a distinct domain.

Within each microservice, we define domain-specific types that serve as foundational abstractions, characterizing the data manipulated by the service. Building upon these types, we establish database schemas representing the persistent state managed by the microservice. Subsequently, we design and implement APIs, which are externally accessible endpoints, providing explicit interaction points for clients and other microservices. Internally, we also define processes—functions encapsulating common business logic that are not directly exposed externally, but rather utilized by the APIs themselves. All functions handling data manipulation, database interactions, APIs, and internal processes are implemented as pure functional constructs  leveraging Scala’s monadic structures. 

This structured, monadic approach also facilitates formal verification. Specifically, each API function invocation can be systematically decomposed into combinations of smaller, pure functions, including process functions, database table operations, and inter-service API calls. Each individual function, due to its purity and well-defined typing, can be translated directly into Lean.

\subsection{Lean 4 and Dependent Types}
\label{sec:lean4_and_dependent_types}

Lean 4~\citep{moura2021lean} is a functional programming language and proof assistant designed specifically for formal verification, featuring a powerful and expressive type system. Its syntax and functional programming style bear similarities to Scala.

The key advantage of Lean 4 lies in its support for \textbf{dependent types}, significantly enhancing its expressiveness and verification capabilities. Dependent types allow types to be parameterized by values, enabling the precise encoding of program invariants directly into the type system. For instance, a dependent type can express properties such as the length of a list being exactly \( n \), or an integer being within certain bounds.

This mechanism tightly couples code with properties, enabling formal verification of program behavior directly through type checking. Lean's dependent type system thus provides a rigorous bridge between Scala’s functional implementations and formal proofs, effectively supporting our verification pipeline.

\section{Method}
\label{sec:method}

\subsection{Pipeline Overview}
\label{sec:pipeline_overview}

\begin{figure}[htbp]
    \centering
    \includegraphics[width=0.9\linewidth]{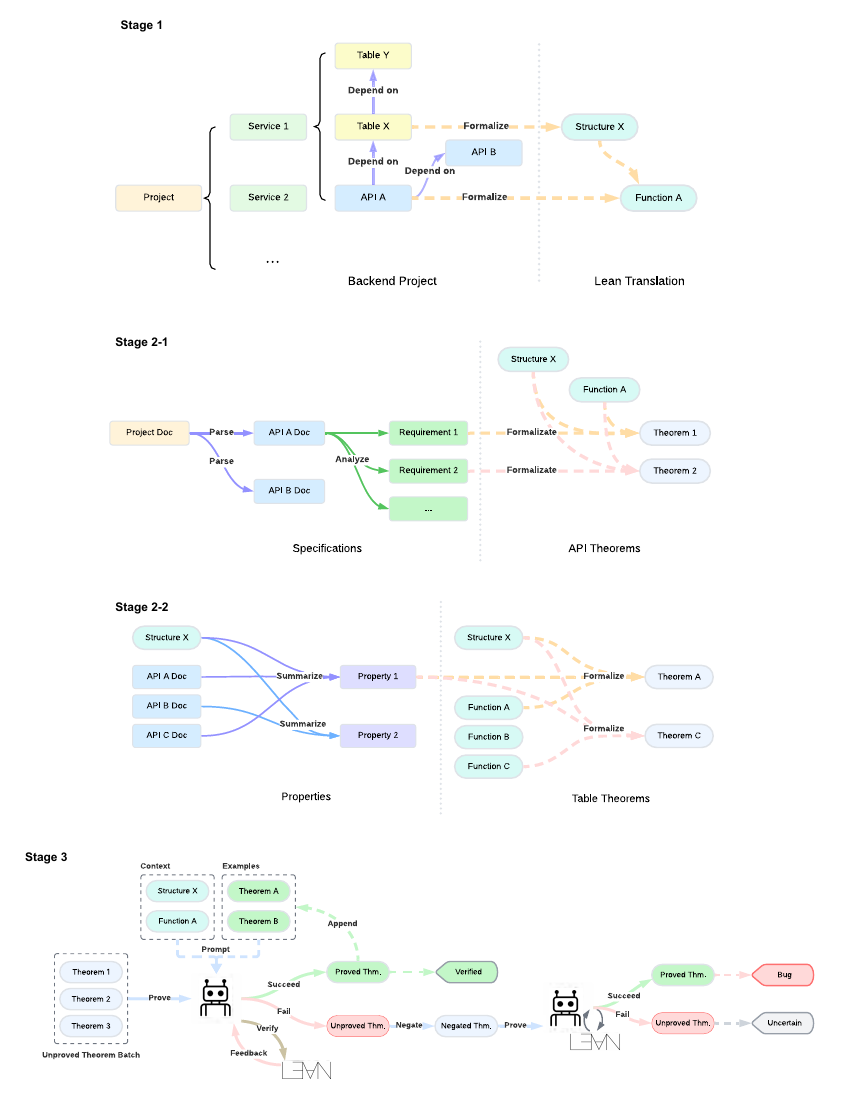}
    \caption{\textbf{Overview of the verification pipeline.} \textbf{Stage 1:} the implementation formalization stage, including the dependency analysis and the formalization process of table structures and API functions.  \textbf{Stage 2-1:} API theorem generation, including the decomposition of system specifications into API requirements, and the construction of formal API theorems based on the API and its dependencies. 
  \textbf{Stage 2-2:} table theorem generation, including the summarization of table properties derived from a subset of related APIs (A and C for property 1 here), and these properties are used to generate formal theorems linked to the corresponding APIs.  \textbf{Stage 3:} proof search stage, where an LLM prover attempts to prove a batch of unverified theorems. If a theorem is successfully proved, its proof is added to the example set. If a proof fails, the prover refines it based on Lean compiler feedback. Theorems that remain unprovable after a set number of refinement rounds are converted into negative theorems that attempt to prove the opposite statement. If a negative theorem is successfully proved, it indicates a bug in the system. Both API theorems and table theorems follow this iterative proving approach.}
    \label{fig:pipeline_stages}
\end{figure}

This section provides a high-level overview of our verification pipeline, which is shown in Fig~\ref{fig:pipeline_stages}, with detailed explanations following in subsequent sections. The input of the pipeline contains two primary components: the backend codebase, which follows the structure outlined in Section~\ref{sec:backend_system_structure}, and a documentation describing the expected API behaviors and their interactions, which serves as the guideline for the verification process.
Given these two components, the pipeline verifies whether the implementation matches the specifications through three stages:

\begin{enumerate}
    \item \textbf{Implementation Formalization:} Given the codebase without documentation, LLMs transform the backend codebase into a Lean 4 project, mapping services, APIs, and database schemas into structured formal representations following predefined patterns. Language-specific details are removed and monads used in the programming are replaced following the strategy introduced in Section~\ref{sec:algebraic_effects_and_handlers}, resulting in a purely functional implementation in Lean that treats all components as dependent types.
    
    \item \textbf{Theorem Generation:} Based on the specifications and formalized implementation, LLMs produce theorems to be proved together with intermediate natural language representations. The models additionally infer and formalize system-wide properties about the tables by viewing all the APIs that interact with the tables as a unified system.

    \item \textbf{Proof Search:} The reasoning model conducts proof searches using a few-shot learning approach with Lean compiler-guided refinement. For unprovable theorems, the system automatically tests their negations to identify potential counterexamples, which detects possible bugs.
\end{enumerate}


Through this process, the implementation is fully formalized in Lean 4, and outputs: (1) verified theorems confirming
specification compliance, (2) counterexamples revealing specification violations,
and (3) unresolved theorems requiring manual inspection.

\textbf{Illustrative Example.} The following sections provide detailed explanations for each stage of the pipeline. To better illustrate our pipeline design, we present concrete examples from a real-world \texttt{BankAccount} system backend that supports user management along with withdrawal, deposit, and balance query functionalities along with the introduction to the pipeline.

\subsection{Implementation Formalization}
\label{sec:implementation_formalization}

The first stage aims at translating the backend codebase into a Lean 4 project. APIs and tables are translated into dependent types like pure functions and structures in Lean, while maintaining the logic of the code and dependencies between them. During the formalization, monads are replaced with explicitly defined inductive return types, while tables are passed as input parameters and outputs of API functions to create pure functional formalizations. 
The formalization stage consists of two phases:
\textbf{dependency analysis} and \textbf{formalization}.

\textbf{Dependency Analysis.} As a pre-task of the formalization, given the API code and table documentation, LLMs are responsible for analyzing dependencies among the following components: Table-to-table dependencies, such as foreign key constraints; API-to-table dependencies, including read and write operations; and API-to-API dependencies, where an API calls other APIs within the same or different services.

\textbf{Formalization.} Following dependency analysis, we formalize the tables
and APIs in topological order. For tables, the formalization yields a Lean
structure, and the APIs will become functions in Lean.

Note that database operations appear as side effects in API implementations, in the format of SQL query execution. The LLMs demonstrate the ability to interpret these
SQL queries and translate them into operations on table structures. To formalize
APIs in a purely functional manner, we treat the database state as a function
parameter, requiring each function's output to include the updated database state.
This approach provides a formal representation of database changes resulting
from API execution.

\begin{figure}[htbp]
    \centering
    \includegraphics[width=0.7\linewidth]{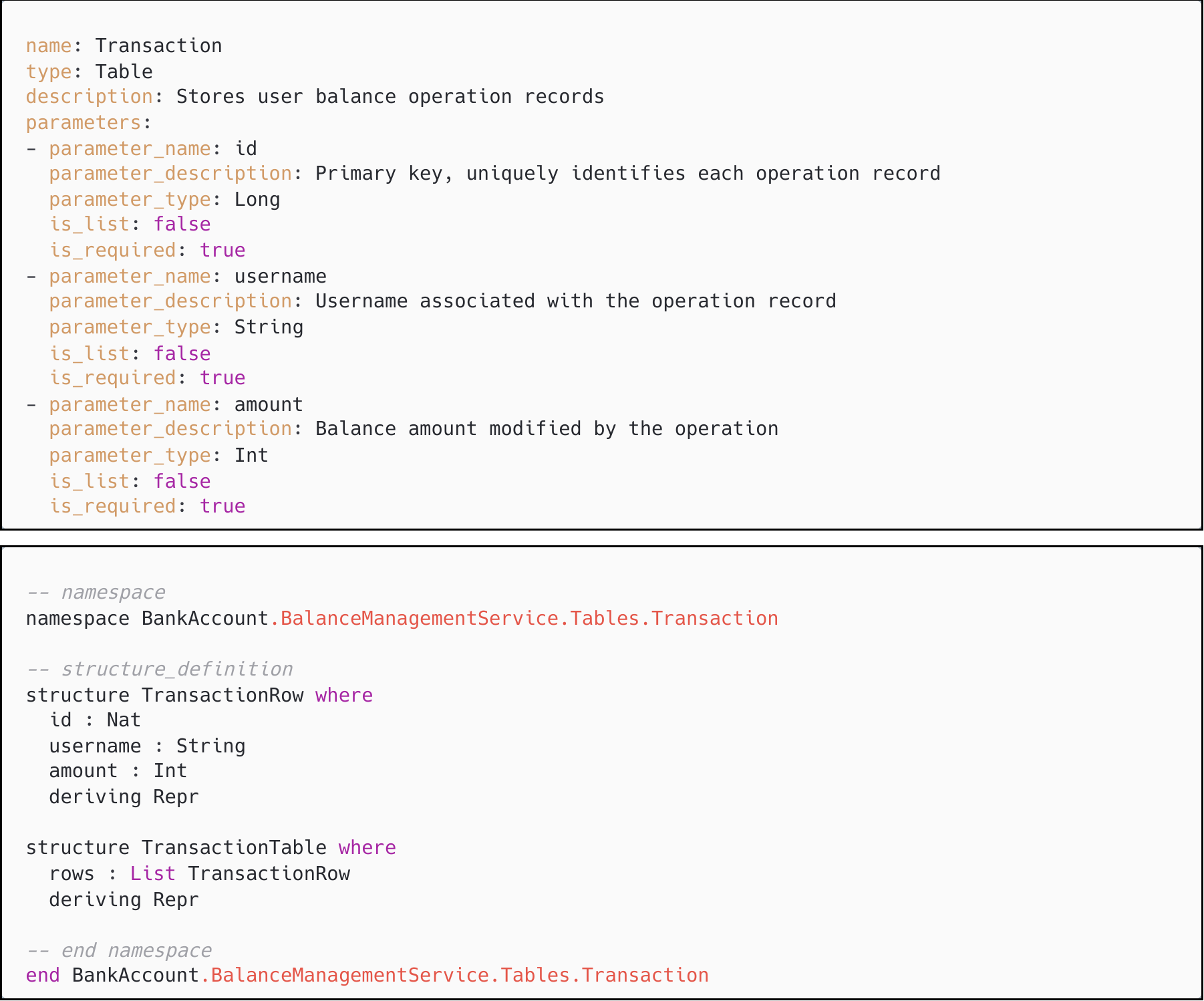}
    \caption{\textbf{An example of formalizing a table structure.} The \texttt{Transaction} table structure is derived from the given table description.}
    \label{fig:table_formalization_example}
\end{figure}

\begin{figure}[htbp]
    \centering
    \includegraphics[width=\linewidth]{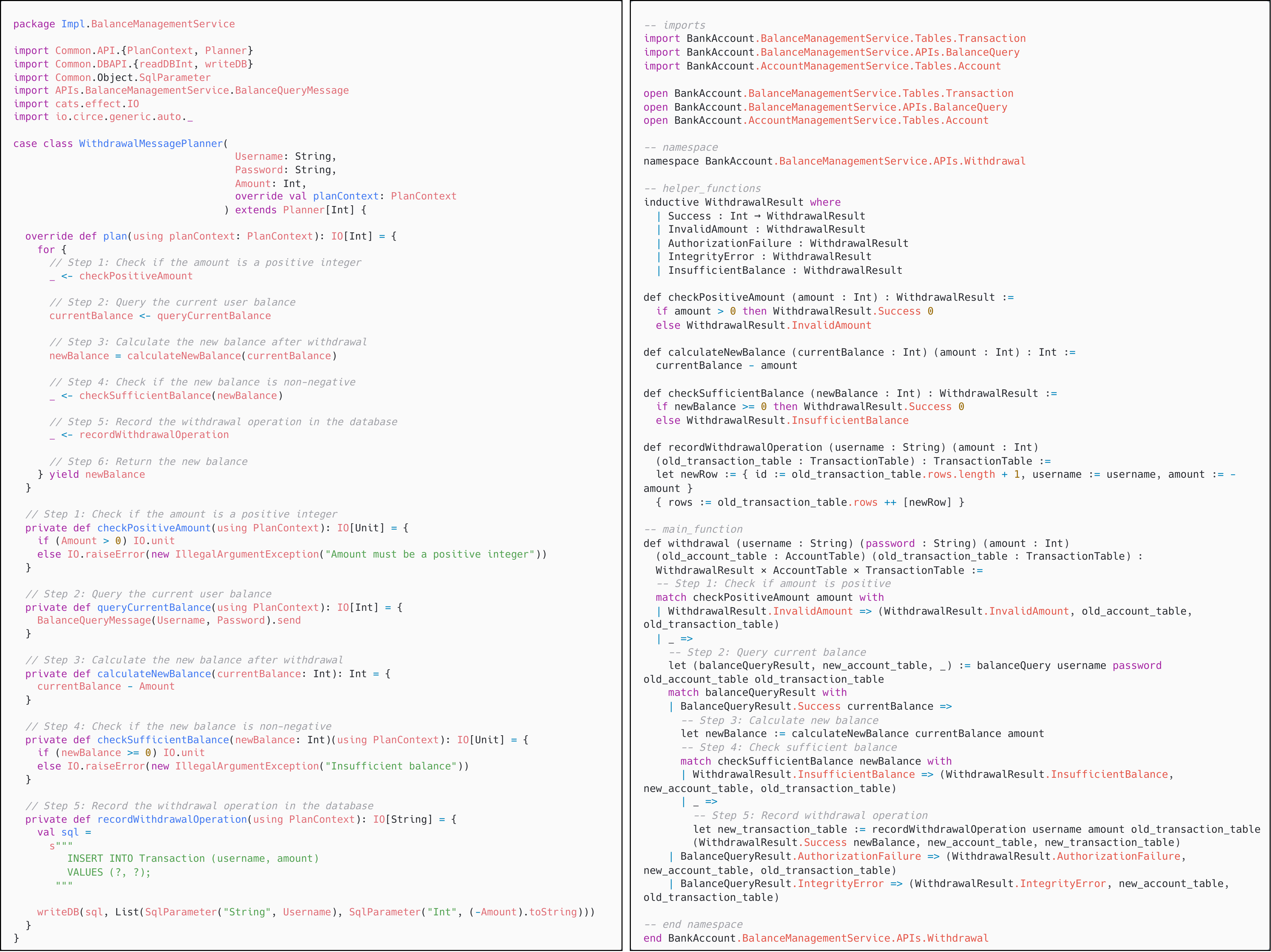}
    \caption{\textbf{An example of formalizing an API in Lean.} The \texttt{WithDrawal} API is formalized based on the implementation code, along with the tables (\texttt{Transaction}) and APIs (\texttt{BalanceQuery}) it depends on.}
    \label{fig:api_formalization_example}
\end{figure}

Fig \ref{fig:table_formalization_example} illustrates table formalization using the \texttt{Transaction} table. The model transforms the table's YAML description into a formal structure containing row and table definitions. Fig \ref{fig:api_formalization_example} demonstrates API formalization. After formalizing all tables, APIs are formalized in topological order: first \texttt{BalanceQuery}, followed by the \texttt{Withdrawal} API that depends on it. The left side shows the \texttt{Withdrawal} implementation, which the model uses alongside formal table representations and dependent APIs to generate the formal function on the right. The formalization process defines return types, including success types with values and various error types, formalizes helper functions following the original logic, and then defines the main function. Since \texttt{BalanceQuery} depends on both \texttt{Account} and \texttt{Transaction} tables, their initial states must be included in the \texttt{Withdrawal} function inputs, with updated states returned.

Upon completion of this stage, the codebase undergoes a formal translation into an
equivalent Lean 4 project.


\subsection{Theorem Generation}
\label{sec:theorem_generation}

Having constructed a Lean project that preserves the semantic equivalence of the original codebase, we proceed to formalize the system specifications as verifiable theorems within Lean. There are two kinds of theorems: those related to the API specifications and those related to table properties. They are all constructed through a two-step approach, first generating the natural language description of the theorem and then formalizing it in Lean.

\subsubsection{API Theorem Generation}
\label{sec:api_theorem_generation}

\textbf{Requirements Generation.} The input to this step is the system specifications document written in natural language. We employ LLMs to generate detailed documentation for each API from this document. In our experiments, we provide structured documentation that precisely defines the specifications of each API, which facilitates accurate evaluation during proof search.


Then the documentation of each API is split into a list of input-output requirements, where each entry specifies the conditions on the inputs and table states, along with the expected outputs and updated tables, representing a control flow path in the current API.

The functional nature of our API formalization enables isolated verification by treating dependent API returns as preconditions, allowing independent requirement generation for each API, which is a significant advantage for efficiency and scalability.

\textbf{Theorem Formalization.} Then, we formalize the generated requirements as Lean theorems,
with each theorem verifying a distinct control flow path of the API. Building upon
the formal implementation of the API and its dependent tables and APIs, LLMs
follow a structured analysis process to identify theorem components: input parameters
as function arguments, conditions as hypotheses, and expected outputs with updated
table states as conclusions. The LLMs then construct well-formed theorems that
undergo compilation to verify syntactic correctness, with iterative refinement addressing
any detected errors.

\begin{figure}[htbp]
    \centering
    \includegraphics[width=0.7\linewidth]{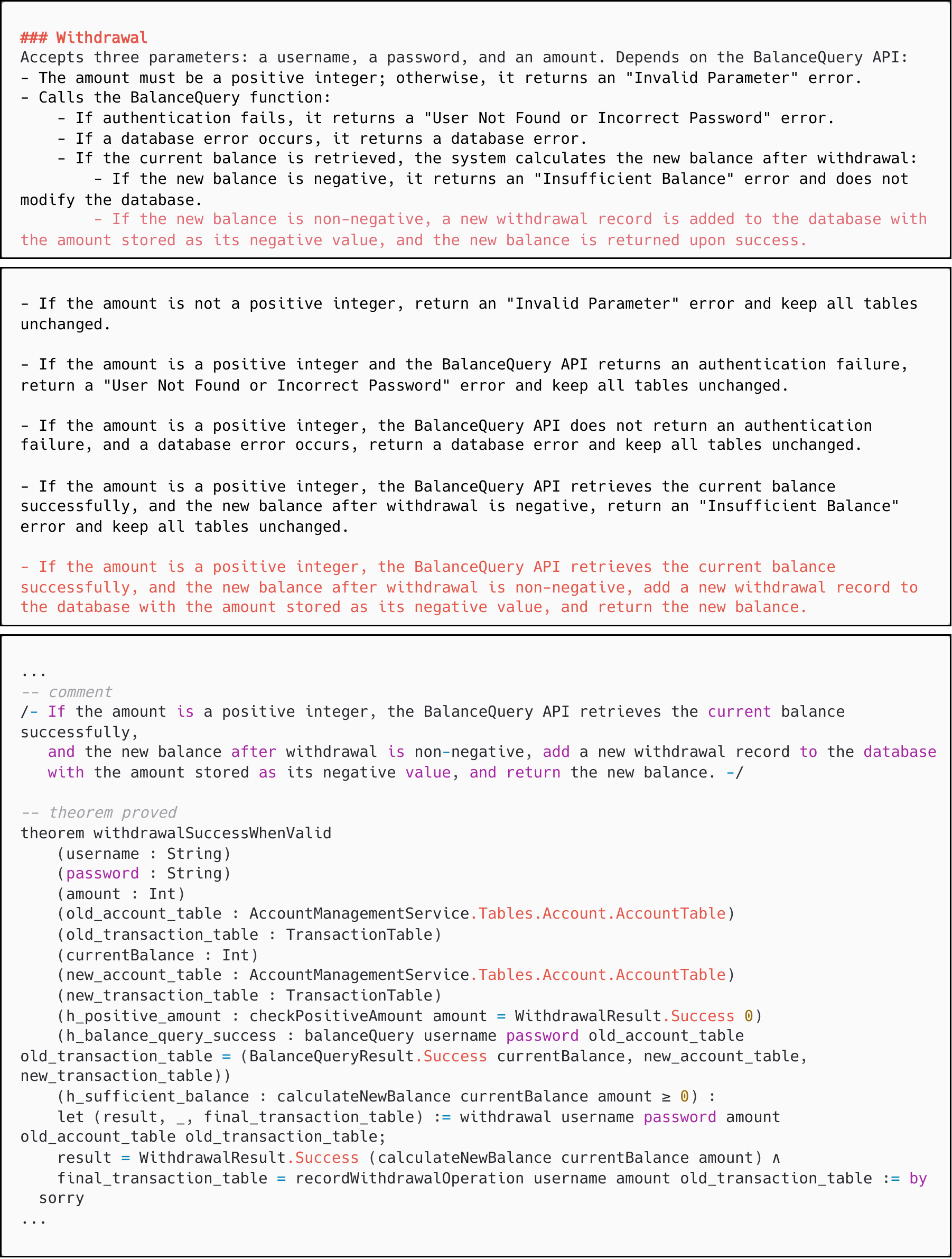}
    \caption{\textbf{An example of API theorem generation following a sequential process: API document interpretation, requirement analysis, and formal theorem formulation.} The document on the top is parsed to produce parallel requirements presented in the middle, with the successful control flow (highlighted in red) becoming the final requirement after necessary premises are added. This requirement is then analyzed to construct the complete formal theorem.}
    \label{fig:api_theorem_example}
\end{figure}

Fig \ref{fig:api_theorem_example} demonstrates an example of API theorem generation. Beginning with the final line of the \texttt{Withdrawal} document parsed from the system specifications, which specifies success conditions and results of the API, the model identifies necessary premises such as positive withdrawal amount, successful \texttt{BalanceQuery} execution, and sufficient balance to create a self-contained requirement. This requirement is then translated into a rigorous formal theorem in Lean. Standard prefixes like imports and open commands are omitted. Before the theorem statement, a comment derived from the requirement explains the theorem's meaning. The theorem statement declares API function input parameters as variables, along with initial states of related tables. All premises from the requirement appear as hypotheses, while the API output and updated table states constitute the conclusion. Currently, the proof is omitted and replaced with a \texttt{sorry} placeholder.

\subsubsection{Table Theorem Generation}

The approach of generating table theorems is similar to that of API theorems, yet we focus on higher-level properties of the tables that are valid in a system-wide manner, by systematically analyzing the API specifications.

\textbf{Property Summarization.} For table properties, we intentionally avoid
providing predefined documentation since these properties should be inferable
from API specifications. Our approach proceeds as follows: First, we aggregate all
APIs interacting with a given table. Next, we analyze their specifications to derive
the table's properties. The LLMs generate each property as a natural language
statement accompanied by the set of APIs responsible for maintaining it. For instance,
read-only APIs must not modify the table.

\textbf{Theorem Formalization.} Each property is formalized as a set of theorems, with one theorem per associated API. These theorems resemble API theorems but have two key differences: they place fewer restrictions on input parameters and concentrate mainly on how tables change rather than on API responses. This approach allows for checking invariants by analyzing how individual API calls modify table states in a single step.

\begin{figure}[htbp]
    \centering
    \includegraphics[width=0.7\linewidth]{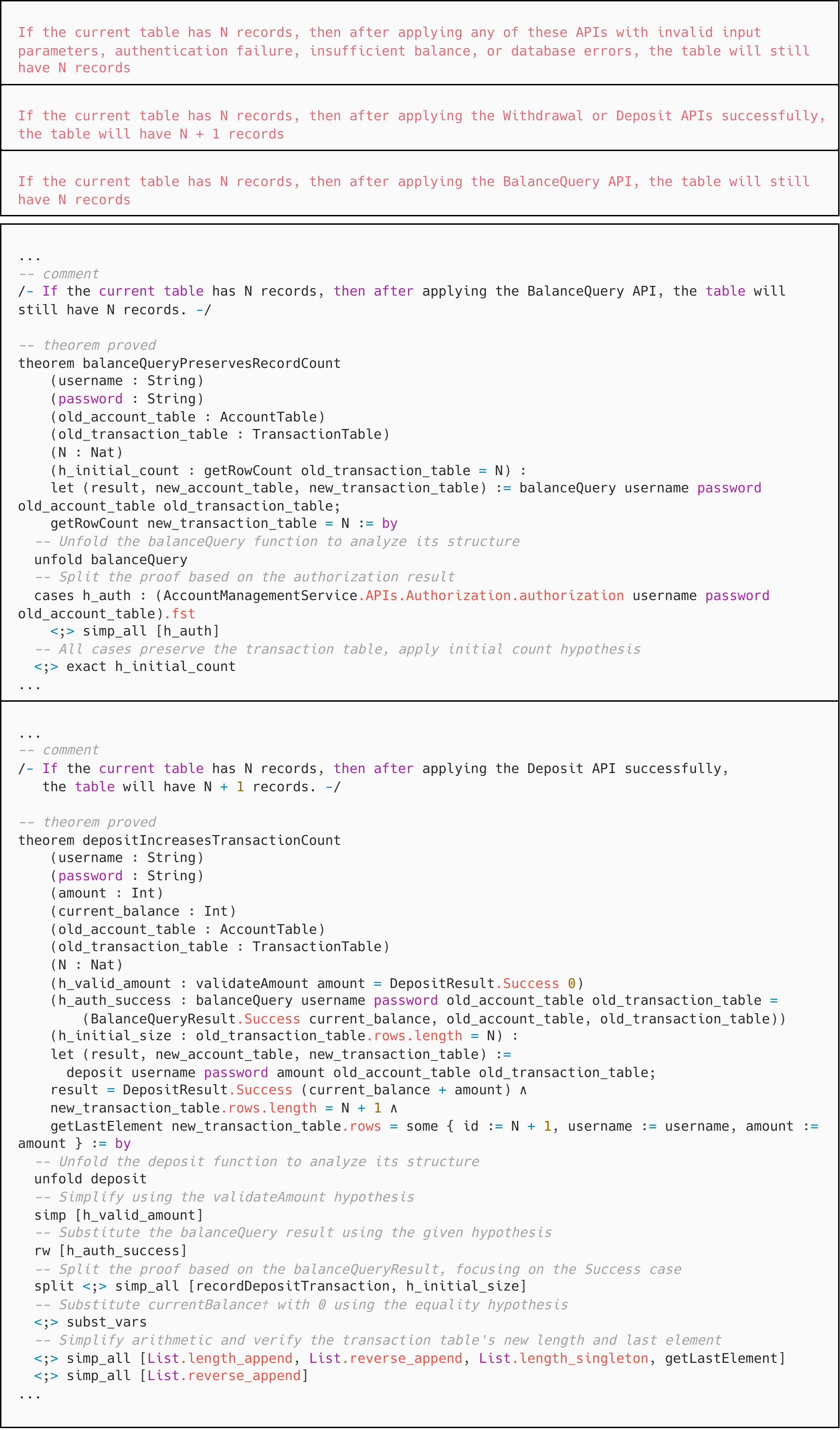}
    \caption{\textbf{Examples of table properties and theorems.} The figure illustrates three properties concerning record numbers in the \texttt{Transaction} table across its three interacting APIs. The first theorem is related to \texttt{BalanceQuery} and is derived from the third property. The second one is associated with the second property, specifically addressing the \texttt{Deposit} operation.}
    \label{fig:table_theorem_example}
\end{figure}

Fig \ref{fig:table_theorem_example} illustrates three properties regarding record quantities in the \texttt{Transaction} table. Three APIs interact with this table: the read-only \texttt{BalanceQuery} never alters records regardless of its outcome, while \texttt{Withdrawal} and \texttt{Deposit} add a new transaction record upon success or maintain the current count if unsuccessful. The theorems derived from these properties address all possible scenarios, with two shown as examples. The \texttt{BalanceQuery} theorem requires no hypotheses about API inputs or success conditions, asserting that the record count remains unchanged under all circumstances. The \texttt{Deposit} theorem, conversely, includes hypotheses ensuring API success, leading to the conclusion that the record count increases by one. Compared to API theorems, these table theorems pay more attention to the table states instead of API outputs, collectively characterizing the properties of the tables in the backend system.

\subsection{Proof Search}
\label{sec:proof_search}

\subsubsection{Proof Strategy}

The final stage of our pipeline employs general LLMs with advanced reasoning
capabilities (e.g., DeepSeek-R1~\citep{guo2025deepseek}) to conduct proof search for theorems concerning
APIs and tables. Our approach incorporates two key strategies:

\textbf{Compiler-Guided Refinement.} When proofs fail, we retrieve compiler error
messages along with their contextual locations, and also implement a
backtracking mechanism to identify the last correct proof step with the
unsolved goal, enabling iterative, step-by-step proof refinement until
successful verification.

\textbf{Adaptive Few-Shot Learning.} For each batch of theorems, we dynamically
select relevant proved theorems based on three levels of similarity: (1) same API/table,
(2) same service, and (3) project-wide. This hierarchical retrieval provides
increasingly relevant proof examples to guide the model. Based on this, we implement a dual-loop
architecture where local theorem refinement coexists with a global retry mechanism—unproved
theorems gain increasing proof opportunities as the pool of verified examples expands
throughout the project.

Each theorem successfully verified during proof search corresponds to a validated control flow path within the API implementation. Examples of proved theorems can be found in Fig~\ref{fig:table_theorem_example} of the table theorem section.

\subsubsection{Counterexample Search}
For those theorems unable to be proved after the entire proof search loop, we employ LLMs
to search for counterexamples by first negating the original
theorem and attempting to prove its negation. This process either produces a verified
counterexample that detects a bug or results
in an undetermined case requiring human verification.

\begin{figure}[htbp]
    \centering
    \includegraphics[width=\linewidth]{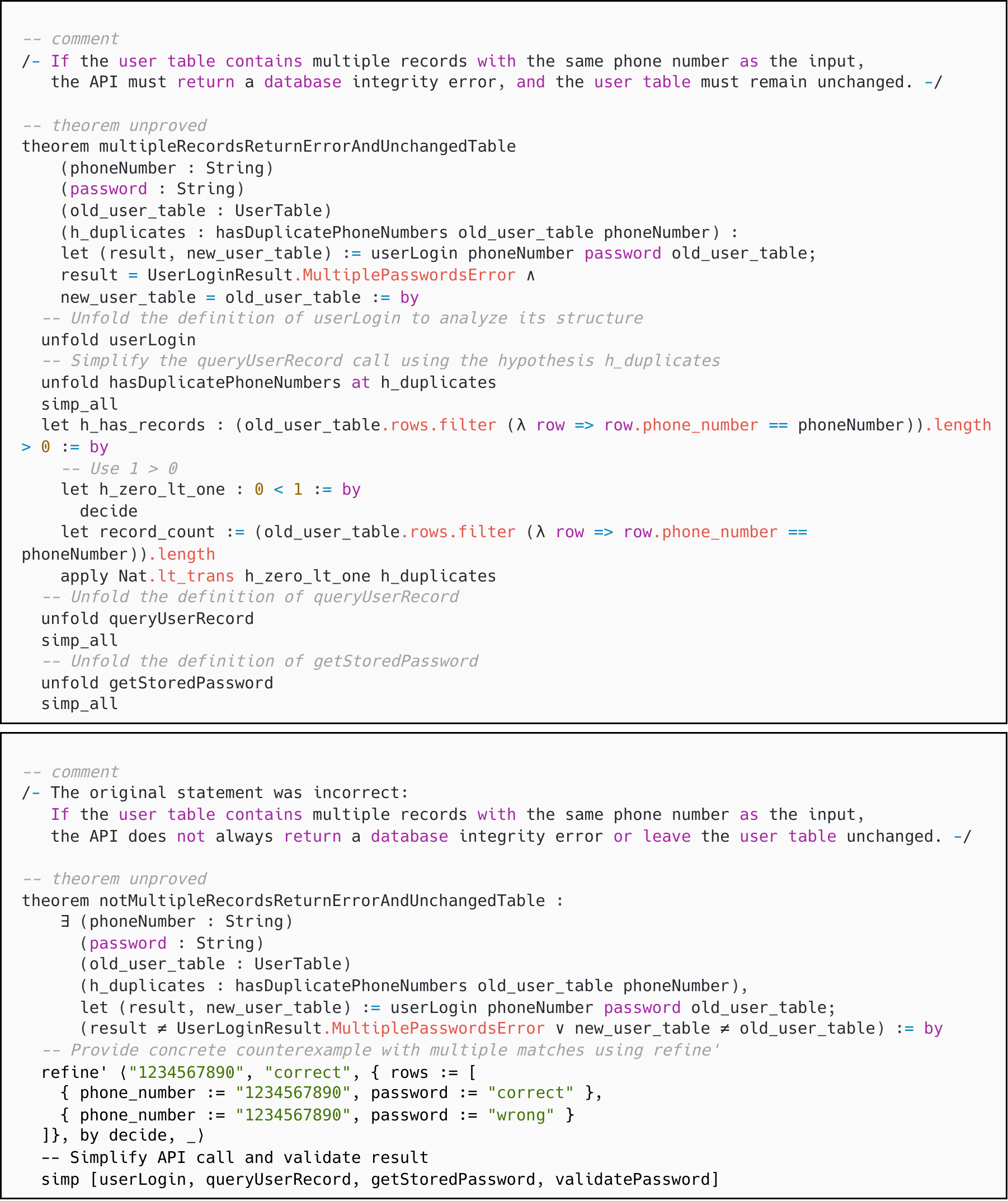}
    \caption{\textbf{An example of negative theorem generation.} The second theorem represents the negated version of the first theorem. Their respective proofs derive from different API implementations—the correct implementation supports the original theorem, while a flawed implementation substantiates the negative theorem.}    \label{fig:negative_theorem_example}
\end{figure}

Fig \ref{fig:negative_theorem_example} demonstrates the example of an original theorem and its negated counterpart. Using the straightforward \texttt{UserLogin} API as an illustration of negative theorem logic, we see how this API verifies whether the table contains multiple records with identical phone numbers. The original theorem presents the duplicated phone number scenario as a hypothesis, asserting that the output will be an error and the table will remain unchanged. In the negated theorem, the objective becomes finding input combinations that satisfy the hypothesis yet contradict the conclusion. Naturally, the proof strategies for these theorems diverge: the original positive theorem requires expanding the API function and its helper functions, then leveraging hypotheses to match cases until arriving at the desired output. In contrast, the negative theorem merely needs a counterexample that satisfies all hypotheses yet contradicts the expected conclusion. It's important to note that the proofs of these two theorems are based on distinct formal implementations of the \texttt{UserLogin} API, as only one theorem can be valid for any given implementation.


\section{Results}
\label{sec:results}

\subsection{Experiment Setup}
\label{sec:experiment_setup}

We evaluate our approach on a dataset comprising four Scala projects developed
within a functional programming framework. These projects vary in size and complexity,
each containing 2 to 3 manually injected bug variants. The projects were generated using
an AI-based project generator to ensure well-structured code organization and were
subsequently verified by human engineers. All projects address real-world needs,
with brief descriptions presented in Table~\ref{tab:project_description}. The
projects contain between 2 and 25 APIs each, yielding more than 150 theorems about
API functionality and 100 theorems about table properties, thus providing a
diverse testbed for evaluation. Detailed project information is shown in Table~\ref{tab:projects}.

\begin{table*}
    [ht]
    \centering
    \caption{\textbf{Project Information for the Scala Projects, including the number of
    total services, tables, APIs, and the number of bug variants.}}
    \label{tab:projects} \small \resizebox{0.55\textwidth}{!}{
    \begin{tabular}{@{}c|c|c|c|c@{}}
        \toprule \textbf{Project} & \texttt{\#}Services & \texttt{\#}Table & \texttt{\#}API & \texttt{\#}Variant \\
        \midrule UserAuth         & 1                   & 1                & 2              & 2                  \\
        BankAccount               & 3                   & 2                & 5              & 3                  \\
        Email                     & 3                   & 3                & 13             & 3                  \\
        TaxiApp                   & 5                   & 3                & 25             & 3                  \\
        \bottomrule
    \end{tabular}
    }
\end{table*}

For our pipeline implementation, we employ qwen-max~\citep{yang2024qwen2} for all tasks except theorem
proving, leveraging its state-of-the-art general capabilities and strong instruction-following
performance. The proving tasks are handled by DeepSeek-R1, which we selected for
its superior reasoning capabilities in formal verification scenarios.

\subsection{Verification Statistics}
\label{sec:verification_statistics}

In this section, we evaluate our pipeline's verification capability using the bug-free
projects in our dataset. To assess each pipeline stage independently, we manually
validate stage inputs to eliminate potential misalignments, which means we use
the best outcome out of three attempts from the last step as the input to the
next step. For non-proving tasks, we execute the pipeline three times to verify
result consistency. The numbers presented in the tables are averaged over the three
times. The theorem proving stage, however, benefits from built-in robustness
through its global retry and local refinement mechanisms, making additional runs
unnecessary—we therefore execute it only once per project.

\subsubsection{Implementation Formalization}

In the implementation formalization stage, there are two main tasks: dependency
analysis and code formalization for tables and APIs. To evaluate dependency analysis,
we measure the number of correctly identified dependencies. For table formalization,
we classify the outcomes into three categories: success, semantic mismatch, and syntax
error. For API formalization, we further refine the categorization into success,
return type mismatch, logic mismatch, and syntax error. The results are presented
in Table~\ref{tab:formalization_results}.

\begin{table*}
    [ht]
    \centering
    \caption{\textbf{Implementation Formalization Results across the Scala projects.} Table
    Dep. and API Dep. refer to dependency analysis for tables and APIs respectively,
    while Table Form. and API Form. refer to formalization results. For all tasks,
    Pos. represents successful cases without any misalignments or errors. For
    dependency analysis, Neg.: Wrong dependencies, Acc\%: Accuracy percentage. For
    Table Formalization, Sem.: Semantic Mismatch, Syn.: Syntax Error. For API
    Formalization, Ret.: Return Type Mismatch, Logic: Logic Mismatch, Syn.:
    Syntax Error.}
    \label{tab:formalization_results} \small \resizebox{\textwidth}{!}{
    \begin{tabular}{@{}c|ccc|cccc|ccc|ccccc@{}}
        \toprule \multirow{2}{*}{\textbf{Project}} & \multicolumn{3}{c|}{\textbf{Table Dep.}} & \multicolumn{4}{c|}{\textbf{Table Form.}} & \multicolumn{3}{c|}{\textbf{API Dep.}} & \multicolumn{5}{c@{}}{\textbf{API Form.}} \\
                                                   & Pos.                                     & Neg.                                      & Acc\%                                  & Pos.                                     & Sem. & Syn. & Acc\% & Pos. & Neg. & Acc\% & Pos. & Ret. & Logic & Syn. & Acc\% \\
        \midrule UserAuth                          & 1.0                                      & 0                                         & 100.0                                  & 1.0                                      & 0    & 0    & 100.0 & 2.0  & 0    & 100.0 & 2.0  & 0    & 0     & 0    & 100.0 \\
        BankAccount                                & 2.0                                      & 0                                         & 100.0                                  & 2.0                                      & 0    & 0    & 100.0 & 5.0  & 0    & 100.0 & 5.0  & 0    & 0     & 0    & 100.0 \\
        Email                                      & 3.0                                      & 0                                         & 100.0                                  & 3.0                                      & 0    & 0    & 100.0 & 13.0 & 0    & 100.0 & 12.3 & 0    & 0.7   & 0    & 94.6  \\
        TaxiApp                                    & 3.0                                      & 0                                         & 100.0                                  & 3.0                                      & 0    & 0    & 100.0 & 25.0 & 0    & 100.0 & 24.0 & 0.7  & 0.3   & 0    & 96.0  \\
        \midrule \textbf{Total}                    & 9.0                                      & 0                                         & 100.0                                  & 9.0                                      & 0    & 0    & 100.0 & 45.0 & 0    & 100.0 & 43.3 & 0.7  & 1.0   & 0    & 96.2  \\
        \bottomrule
    \end{tabular}
    }
\end{table*}

Our results demonstrate that the model achieves perfect accuracy on both dependency analysis and table formalization tasks. For API formalization, despite a small probability of misalignment, the majority of APIs are correctly formalized. Notably, return type misalignments rarely impact the verification process, as the model typically only adds extra return types rather than omitting essential ones.

\subsubsection{Theorem Generation}
\label{sec:theorem_generation_result}

In the theorem generation stage, parsing API documentation from the given backend system specification is relatively straightforward, so we don't treat it as a separate evaluation task. Instead, we evaluate four key tasks: API requirement generation, API theorem formalization, table property summarization, and table theorem formalization. For the natural language-based tasks, we assess the proportion of concise expressions that correctly describe input conditions and output types. For API requirements, we also verify the completeness of control flow paths. Additionally, for table property summarization, we confirm whether relevant APIs are correctly identified. For theorem formalization, results are classified into three categories: success, semantic mismatch, and syntax error. Table~\ref{tab:theorem_generation_results} presents these findings.

\begin{table*}
    [ht]
    \centering
    \caption{\textbf{Theorem Generation Results across the Scala projects.} API Req. and
    Tab. Prop. refer to API requirement generation and table property
    summarization respectively, while API Theorem and Tab. Theorem refer to
    theorem formalization results. For natural language tasks, Pos. represents successful
    cases with concise expressions, Neg.: cases with verbose or incorrect
    expressions. For API requirement generation, Mis.: missing any control flow
    path of the API. For table property summarization, API Err.: incorrect
    identification of relevant APIs. For theorem formalization, Sem.: Semantic
    Mismatch, Syn.: Syntax Error.}
    \label{tab:theorem_generation_results} \small \resizebox{\textwidth}{!}{
    \begin{tabular}{@{}c|cccc|cccc|cccc|cccc@{}}
        \toprule \multirow{2}{*}{\textbf{Project}} & \multicolumn{4}{c|}{\textbf{API Req.}} & \multicolumn{4}{c|}{\textbf{API Theorem}} & \multicolumn{4}{c|}{\textbf{Tab. Prop.}} & \multicolumn{4}{c@{}}{\textbf{Tab. Theorem}} \\
                                                   & Pos.                                   & Neg.                                      & Mis.                                     & Acc\%                                       & Pos.  & Sem. & Syn. & Acc\% & Pos. & Neg. & API Err. & Acc\% & Pos. & Sem. & Syn. & Acc\% \\
        \midrule UserAuth                          & 7.0                                    & 0                                         & 0                                        & 100.0                                       & 7     & 0    & 0    & 100.0 & 6.7  & 0    & 0        & 100.0 & 7.7  & 0    & 0.3  & 95.8  \\
        BankAccount                                & 19.3                                   & 0                                         & 0                                        & 100.0                                       & 17    & 0    & 1    & 94.4  & 12.7 & 0    & 0        & 100.0 & 16.3 & 2.3  & 1.3  & 81.7  \\
        Email                                      & 43.3                                   & 0                                         & 0                                        & 100.0                                       & 42.3  & 0    & 0.7  & 98.4  & 21.3 & 0.3  & 0        & 98.5  & 32.0 & 5.3  & 0.7  & 84.2  \\
        TaxiApp                                    & 105.7                                  & 0                                         & 0.3                                      & 99.7                                        & 95.3  & 5.0  & 1.7  & 93.5  & 20.0 & 3.7  & 0        & 84.5  & 56.3 & 5.7  & 2.0  & 88.0  \\
        \midrule \textbf{Total}                    & 175.3                                  & 0                                         & 0.3                                      & 99.8                                        & 161.6 & 5.0  & 3.4  & 95.1  & 60.7 & 4.0  & 0        & 93.8     & 112.3    & 13.3    & 4.3    & 86.5     \\
        \bottomrule
    \end{tabular}
    }
\end{table*}
To guarantee accurate evaluation, we apply comprehensive manual checks on all theorems and related descriptions. For natural language tasks, the model performs surprisingly well on API requirement generation, with all generated requirements being concise. This indicates the model successfully captures all premises and outcomes across each API control flow, with only one missing path.

Unlike API requirements, the model receives no predefined specifications when summarizing table properties. Consequently, some properties are not actually valid, though the related APIs are all correctly identified. These errors in table properties are not too problematic since they serve only as informal hypotheses that will undergo verification in the formal system later.

API theorem formalization achieves an accuracy about 95\%, while table theorem formalization reaches approximately 85\% accuracy. Among the two types of errors, syntax errors prevent certain control flows from being formalized into theorems, which need manual checks. Semantic mismatches produce compilable but requirement-inconsistent theorems, which undermine verification integrity. Most errors in table theorem formalization stem from semantic mismatches caused by the model's misunderstanding of table properties, as table theorems typically describe general properties while the model tends to add extra input conditions rather than providing general theorems, which makes the formalized theorem simpler and not aligned with the description.

\subsubsection{Proof Search}
\label{sec:proof_search_results}

In the proof search stage, we evaluate the pipeline's ability to prove the theorems
generated in the previous stage. Since negated theorems are only relevant when
identifying bugs in the implementation, we do not assess the generation and proof
of negative theorems here, leaving that evaluation to the bug detection section described in Section~\ref{sec:bug_detection}. The
proof search task has only two possible outcomes: success or failure. The results
are presented in Table~\ref{tab:proof_search_results}.

\begin{table*}
    [ht]
    \centering
    \caption{\textbf{Proof Search results across the Scala projects.} The table shows
    the number of theorems successfully proved, those that remained unproved, and
    the overall proving ratio for APIs and tables.}
    \label{tab:proof_search_results} \small \resizebox{0.8\textwidth}{!}{
    \begin{tabular}{@{}c|ccc|ccc@{}}
        \toprule \multirow{2}{*}{\textbf{Project}} & \multicolumn{3}{c|}{\textbf{API Proof}} & \multicolumn{3}{c}{\textbf{Table Proof}} \\
                                                   & Proved                                  & Unproved                                & Proved\% & Proved & Unproved & Proved\% \\
        \midrule UserAuth                          & 4                                       & 3                                       & 57.1        & 3      & 4        & 42.9        \\
        BankAccount                                & 10                                       & 7                                       & 58.8        & 11      & 7        & 61.1        \\
        Email                                      & 34                                       & 8                                       & 81.0        & 24      & 13        & 64.9        \\
        TaxiApp                                    & 52                                       & 50                                       & 51.0        & 34      & 29        & 54.0        \\
        \midrule \textbf{Total}                    & 100                                       & 68                                       & 59.5        & 72      & 53        & 57.6        \\
        \bottomrule
    \end{tabular}
    }
\end{table*}

The results indicate that over 50\% of theorems are successfully proved, enabling verification of half of all API specifications in addition to automatically analyzed and proved table properties, which can reduce manual verification labor costs by half, even though the successfully proven theorems are sometimes relatively simple compared to the unproven ones.

Since no specialized prover model exists for generating proofs for this kind of theorems that is related to real-world logic and has complex outside dependencies, the current results from our general reasoning model are quite promising. These results can be further enhanced by collecting proof data and fine-tuning a task-specific model, suggesting significant potential to both improve accuracy and reduce costs in the proof search stage.

Currently, the proof budget for each theorem allows 5 attempts with 8 refinement retries per attempt. In our evaluation of the TaxiApp project, the average token-based model cost is \$0.34, with verification costs averaging \$2.19 per API, while test engineers in America earn \$52 per hour on average. The expenses from the first two stages are negligible compared to proof costs. These expenses could be reduced by implementing more efficient provers, which would allow for a smaller proof budget.

Notably, we observe that the proving ratio of theorems does not decrease as projects scale in size. This indicates that, given the decomposable nature of our functional programming framework, our pipeline remains scalable to larger projects, while increased coroutine utilization can maintain low time complexity. The consistent accuracy of the theorem generation task in Section~\ref{sec:theorem_generation_result} further supports this conclusion.

\subsection{Bug Detection}
\label{sec:bug_detection}

In this section, we examine the pipeline from a different perspective, treating it
as a unified tool for bug detection. We evaluate its capability by assessing
whether it can identify injected bugs in our dataset across the three stages.

(1) During implementation formalization, the model is allowed to issue warnings about
potential bugs but must still formalize the code exactly as given, even if it
contains errors. (2) In theorem generation, the formalization derived from the buggy
implementation may fail to produce theorems that align with the specifications (e.g.,
missing return types, missing parameters, etc.). The model can flag
inconsistencies between the formalization and the specifications but must still
generate a theorem statement that best matches the intended meaning. (3) Finally, in
the proof search stage, bugs can naturally be detected by identifying counterexamples
when a theorem is not provable.

For evaluation, we check whether a bug is detected at any of these three stages and
count the total occurrences of successful detections. The results are presented
in Table~\ref{tab:bug_detection_results}. For a detailed list of the bugs in the
variants, see Table~\ref{tab:bug_content}.
\begin{table*}
  [ht]
  \centering
  \caption{\textbf{Bug Detection results across the Scala projects.} For each project
  variant, the table shows whether the bug was detected in each of the three
  stages: Impl. Form. (Implementation Formalization), Theorem Gen. (Theorem
  Generation), and Proof Search. The Sum column represents the total number of stages
  that successfully detected the bug (0-3). tick: detected, cross: not detected, circle: unable to detect because of misaligned theorem.}
  \label{tab:bug_detection_results} \small \resizebox{0.8\textwidth}{!}{
  \begin{tabular}{@{}c|c|ccc|c@{}}
    \toprule \textbf{Project}             & \textbf{Variant ID} & \textbf{Impl. Form.} & \textbf{Theorem Gen.} & \textbf{Proof Search} & \textbf{Sum} \\
    \midrule \multirow{2}{*}{UserAuth}    & 1                   & \cmark               & \cmark                & \cmark                & 3            \\
                                          & 2                   & \xmark               & \cmark                & \cmark                & 2            \\
    \midrule \multirow{3}{*}{BankAccount} & 1                   & \xmark               & \xmark                & \cmark                & 1            \\
                                          & 2                   & \xmark               & \cmark                & \cmark                & 2            \\
                                          & 3                   & \xmark               & \xmark                & \xmark                & 0            \\
    \midrule \multirow{3}{*}{Email}       & 1                   & \xmark               & \xmark                & \xmark                & 0            \\
                                          & 2                   & \xmark               & \cmark                & \umark                & 1            \\
                                          & 3                   & \xmark               & \cmark                & \umark                & 1            \\
    \midrule \multirow{3}{*}{TaxiApp}     & 1                   & \xmark               & \cmark                & \umark                & 1            \\
                                          & 2                   & \xmark               & \cmark                & \cmark                & 2            \\
                                          & 3                   & \xmark               & \xmark                & \xmark                & 0            \\
    \bottomrule
  \end{tabular}
  }
\end{table*}

The experiments show that 8 out of 11 bugs are successfully detected by finding counterexamples, while the remaining 3 bugs are still hidden among the theorems that cannot be proved correct or incorrect. According to the results in Section~\ref{sec:proof_search_results}, that proportion is no more than half of all the theorems. 
Among the results, circle mark indicates that this bug prevents the model from writing a theorem aligned with the requirement, so it will be detected in the first two stages and the theorem in the proof search no longer makes sense.

\section{Discussion}
\label{sec:discussion}

We have proposed a novel framework that leverages functional programming and type systems to translate Scala backend code into formal Lean representations, addressing limitations in existing automated testing approaches and expanding automatic testing to system-level backends. By combining LLMs' general capabilities with formal system rigor, our pipeline automatically generates theorems that specify API behavior based on documentation, summarizes and generates table-related properties, and applies LLM-based provers to either verify them through formal proof or detect bugs by finding counterexamples. This approach eliminates the need for human intervention on formally verified functionalities and detected bugs, leaving only the unresolved portion for manual verification.

According to evaluation results on realistic backend systems, our method formally verifies over 50\% of requirements and detects 70\% of bugs in fault variants, automating more than half of the testing workload. The average cost to verify an API is only \$2.19, which is remarkably cost-effective compared to a test engineer's hourly earnings of approximately \$52.

Moreover, our method achieves scalability through the decomposable nature of the functional programming framework. Experimental results in Section~\ref{sec:theorem_generation_result} and ~\ref{sec:proof_search_results} demonstrate that theorem generation accuracy and proof success rates remain consistent regardless of backend size. Since proof search constitutes the majority of pipeline time and cost, verification time complexity for complex projects can be reduced to a constant by simply increasing parallel LLM request executions. This approach offers flexibility comparable to temporarily scaling a testing team—analogous to hiring ten times more test engineers for a project of tenfold complexity and releasing them afterward.

Compared to existing methods focused on automated testing, our approach emphasizes whole-system verification of real-world systems, targeting specification compliance rather than low-level or function-level vulnerabilities. The general capabilities of LLMs make our pipeline scalable not only in project complexity but also across input languages, provided the codebase adheres to functional programming paradigms that enable natural decomposition.

Although our current implementation is based on Scala, the benefits of this approach extend far beyond Scala projects. Scala runs on the JVM and is fully interoperable with Java, allowing it to use any Java library. In particular, Java-based frameworks such as Spring Boot can be systematically translated into Scala, enabling our method to be applied to a wide range of real-world applications that rely on Java ecosystems.

An important requirement of our method is guaranteeing accuracy in two key stages: (1) implementation formalization, ensuring semantic equivalence between implementation and formal representation, and (2) theorem generation, producing formal theorems aligned with natural language specifications. While both stages currently achieve over 95\% accuracy, the remaining misalignments could be further reduced through better specialized models. Additionally, pre-formalizing external services like databases as standard components in the future would help ensure greater equivalence between implementation and formalization.

In summary, our results demonstrate the viability of employing LLMs throughout the entire reliable software testing process, pointing to a promising direction for scalable, AI-powered software testing, with the potential to greatly improve engineering productivity as models continue to advance.













\newpage
\begin{appendices}

\section{Experimental Details}
\label{sec:detailed_result}

\subsection{Data}

\subsubsection{Evaluation Projects}

\begin{table*}[ht]
    \centering
    \caption{\textbf{Brief descriptions of the Scala projects used for evaluation.}}
    \label{tab:project_description} \small \resizebox{\textwidth}{!}{
    \begin{tabular}{@{}c|l@{}}
        \toprule \textbf{Project} & \textbf{Description} \\
        \midrule 
        UserAuth         & Simple user authentication system with login and registration APIs. \\
        BankAccount      & Banking system with APIs for account management and transactions, \\
                        & including deposits and withdrawals. \\
        Email            & Email service for sending and receiving messages, with authentication and group management support. \\
        TaxiApp          & Taxi booking system with account management, ride status tracking controlled by passengers and drivers, \\
                        & and payment processing. \\
        \bottomrule
    \end{tabular}
    }
\end{table*}

\subsubsection{Bug Variants}

\begin{table*}
  [ht]
  \centering
  \caption{\textbf{Descriptions of the bug variants.}}
  \label{tab:bug_content} \small \resizebox{1.0\textwidth}{!}{
  \begin{tabular}{@{}c|c|l@{}}
    \toprule \textbf{Project}             & \textbf{Variant ID} & \textbf{Description}                              \\
    \midrule \multirow{2}{*}{UserAuth}    & 1                   & \texttt{UserRegister} adds record without checking existence.\\
                                          & 2                   & \texttt{UserLogin} has no check for db error when duplicated phone numbers exist.             \\
    \midrule \multirow{3}{*}{BankAccount} & 1                   & \texttt{BalanceQuery} sums all transactions instead of those of the user. \\
                                          & 2                   & \texttt{Deposit} doesn't check positive amount.             \\
                                          & 3                   & \texttt{Withdrawal} stores (amount) in table instead of (-amount)                              \\
    \midrule \multirow{3}{*}{Email}       & 1                   & \texttt{AddUser} doesn't check if the current user is the owner of the group.                      \\
                                          & 2                   & \texttt{DeleteUser} allows the deletion of the owner of the group.       \\
                                          & 3                   & \texttt{SendToGroup} calls the wrong dependent API to check only group exists instead of the user in the group.       \\
    \midrule \multirow{3}{*}{TaxiApp}     & 1                   & \texttt{Login} doesn't delete old tokens when exist.       \\
                                          & 2                   & \texttt{UpdateRideStatus} allows to move from \texttt{awaiting\_payment} to \texttt{cancelled}.              \\
                                          & 3                   & \texttt{PayForRide} only checks if the token is a passenger instead of checking if it matches the \texttt{ride\_id}.                              \\
    \bottomrule
\end{tabular}
  }
\end{table*}




\end{appendices}


\newpage

\bibliography{main}

\end{document}